\documentclass[conference]{IEEEtran}
\IEEEoverridecommandlockouts

\usepackage{siunitx}
\usepackage{cite}
\usepackage{amsmath, amssymb, amsfonts}
\usepackage{algorithmic}
\usepackage{textcomp}
\usepackage{graphicx}
\usepackage{xcolor}
\usepackage{hyperref}
\usepackage{listings}
\usepackage{adjustbox}

\definecolor{pblue}{rgb}{0.13,0.13,1}
\definecolor{pgreen}{rgb}{0,0.5,0}

\renewcommand{\baselinestretch}{0.98}
\def\BibTeX{{\rm B\kern-.05em{\sc i\kern-.025em b}\kern-.08em T\kern-.1667em\lower.7ex\hbox{E}\kern-.125emX}}
\begin{document}
    \bstctlcite{myref:BSTcontrol}
    \title{Accelerating Elliptic Curve Point Additions on Versal AI Engine for
    Multi-scalar Multiplication}

    \author{\IEEEauthorblockN{Ayumi Ohno, Kotaro Shimamura, Shinya Takamaeda-Yamazaki}
    \IEEEauthorblockA{\textit{The University of Tokyo} \\ \{ayumi0130ohno, kotaro, shinya\}@is.s.u-tokyo.ac.jp}
    }

    \maketitle

    \begin{abstract}
        Multi-scalar multiplication (MSM) is crucial in cryptographic applications
        and computationally intensive in zero-knowledge proofs. MSM involves
        accumulating the products of scalars and points on an elliptic curve
        over a 377-bit modulus, and the Pippenger algorithm converts MSM into a series
        of \textit{elliptic curve point additions} (PADDs) with high parallelism.

        This study investigates accelerating MSM on the Versal ACAP platform, an
        emerging hardware that employs a spatial architecture integrating 400 AI
        Engines (AIEs) with programmable logic and a processing system. AIEs are
        SIMD-based VLIW processors capable of performing vector multiply-accumulate
        operations, making them well-suited for multiplication-heavy workloads in
        PADD. Unlike simpler multiplication tasks in previous studies, cryptographic
        computations also require complex operations such as carry propagation.
        These operations necessitate architecture-aware optimizations, including
        \textit{intra-core} dedicated coding style to fully exploit VLIW
        capabilities and \textit{inter-core} strategy for spatial task mapping.

        We propose various optimizations to accelerate PADDs, including (1)
        algorithmic optimizations for carry propagation employing a carry-save-like
        technique to exploit VLIW and SIMD capabilities and (2) a comparison of
        four distinct spatial mappings to enhance intra- and inter-task parallelism.
        Our approach achieves a computational efficiency that utilizes 50.2\% of
        the theoretical memory bandwidth and provides 568$\times$ speedup over
        the integrated CPU on the AIE evaluation board.
    \end{abstract}


    \section{Introduction}
    Multi-scalar multiplication (MSM) is a computationally demanding operation
    in the proof generation process of zk-SNARKs. MSM accumulates the multiplication
    of scalars and points on an elliptic curve, which requires numerous point
    additions (PADDs). The Pippenger algorithm reduces the number of PADD operations
    and enhances parallelism. It partitions points into buckets based on sliced
    scalar values, accumulating points within each bucket.

    Accelerators based on GPUs, FPGAs, and ASICs have been proposed to address the
    computational demands of MSM. GPU-based solutions \cite{ma2023gzkp}
    distribute PADD operations across multiple cores to achieve high parallelism,
    but workload balancing and energy efficiency pose significant challenges. In
    contrast, FPGA-based \cite{zhao2023bstmsm}\cite{aasaraai2022fpga}
    accelerators typically employ a few fully pipelined PADD units to support
    continuous input processing; these solutions rely on sophisticated scheduling
    algorithms to mitigate bucket collisions and stalls.

    Recent advancements in heterogeneous platforms, such as Versal ACAP, offer the
    potential to combine GPU-like computational power with FPGA-like flexibility
    at low power consumption. Versal ACAP integrates AI Engines (AIEs) running
    at high clock frequencies, programmable logic (PL), and a processing
    subsystem (PS) on a single chip. This study explores a novel approach to accelerating
    MSM on Versal AI Engine by leveraging AIEs for compute-intensive PADD operations
    and PL for bucket management. Specifically, we focus on optimizing PADD on AIEs
    and evaluating its performance in the context of MSM.

    Most existing accelerators on Versal AI Engine \cite{zhuang2023charm}\cite{dong2024eq}
    target matrix multiplication or convolutions, which primarily involve simple
    multiply-accumulate (MAC) operations and data transfers. However, the potential
    of AIEs for general-purpose and cryptographic applications remains underexplored.
    AIM \cite{yang2023aim} is one of the few studies that accelerates arbitrary-precision
    integer operations on AIEs. However, it performs carry propagation on the PL,
    leading to significant data transfer overhead for small bit sizes. For MSM
    targeting the BLS12-377 curve, 377-bit integers are too small to benefit from
    AIM’s approach.

    To address the high data transfer overhead caused by performing carry propagation
    outside the AIE, we propose optimizations for effectively handling carry
    propagation and continuous multiplications within the AIE. To mitigate numerous
    scalar operations overhead of carry propagation, we leverage Very Long
    Instruction Word (VLIW) to enable scalar operations to parallel vector MAC
    operations. Additionally, we adopt a carry-save-like technique to minimize
    scalar overhead and further increase parallelism. Finally, we evaluate four
    tiling strategies to enhance VLIW utilization within individual AIE tiles, increase
    parallelism across multiple tiles within the same unit, and improve
    parallelism across different units.

    \section{Background}
    \subsection{Versal AI Engine}
    Versal ACAP \cite{1} integrates PL, PS, and 400 AIEs into a single chip. AIEs
    are SIMD processors running at 1.25 GHz, executing seven-way VLIW
    instructions per cycle. The vector unit supports MAC, addition, shifts, and
    permutation, achieving eight parallel MACs in 32-bit fixed-point mode.

    AIEs are organized in a 50 × 8 array, with each AIE communicating via AXI4
    stream, Cascade stream, and Shared memory. AXI4 stream connects AIEs through
    32-bit switches, limited to six south-to-north and four in other directions;
    packet-switched streams share ports when exceeding these limits. The Cascade
    stream offers higher bandwidth with a 384-bit accumulator but only connects
    neighboring AIEs in the same row. Shared memory allows neighboring AIEs to access
    memory using 256-bit load and store units. AIE and PL interfaces are
    connected via PLIO, providing up to 1.2 TB/s (from PL) and 0.9 TB/s (to PL) at
    500 MHz. The VCK190 includes DDR DRAM with 25.6 GB/s bandwidth per DIMM.
    \subsection{Large Integer Multiplication on Versal AI Engine}
    \label{AA} AIM \cite{yang2023aim} utilizes the schoolbook algorithm to accelerate
    arbitrary-precision integer multiplication on Versal AI Engine. It divides
    multiplications into several blocks and assigns each to an AIE core, while the
    PL handles carry propagation. It enables data reuse across rows and columns
    by assigning multiple sub-blocks to a single core, improving the computation-to-data-transfer
    ratio. AIM targets significantly larger integers than the 377-bit numbers
    used in MSM. It does not achieve throughput gains over CPUs or energy efficiency
    improvements over GPUs for integers smaller than 8096 bits.

    \subsection{Multi-Scalar Multiplication}
    MSM computes the sum of multiple scalars multiplied by corresponding points on
    an elliptic curve. MSM accelerators often employ the Pippenger algorithm, which
    reduces PADD operations and enables parallel processing. The algorithm
    segments $k$-bit scalars into $w$-bit windows and organizes them into $k/w$
    groups. Each group's points are assigned to buckets based on the sliced scalar
    values and accumulated (bucket accumulation). The accumulated buckets are
    then processed iteratively to aggregate the results (bucket aggregation). Finally,
    the aggregated results from all groups are combined to produce the final
    output (group aggregation).

    PADD operations can be pipelined in bucket accumulation if successive
    operations do not have data dependency. 
    Techniques like bucket state management in BSTMSM \cite{zhao2023bstmsm} and
    FIFO utilization in CycloneMSM \cite{aasaraai2022fpga} allow continuous
    point input even when successive points are in the same bucket. In contrast,
    PADD operations must be executed sequentially during bucket and group
    aggregation, waiting for the completion of the previous PADD. This study
    designs separate PADD units on the AIEs tailored to meet the distinct
    requirements. 
    \subsection{Elliptic-curve addition}
    An elliptic curve is defined by the equation $y^{2}= x^{3}+ ax + b$ over a
    finite field $GF(p)$. Similar to other MSM accelerators \cite{zhao2023bstmsm}\cite{aasaraai2022fpga},
    we target the BLS12-377 curve and convert the curve into the Twisted Edwards
    curve representation $- x^{2}+ y^{2}=1 + \frac{k}{2}x^{2}y^{2}$. In addition
    to $(x, y)$ affine coordinates, a new parameter $t = xy$ is introduced and converted
    into projective coordinates $(X, Y, T, Z)$, where $X = x \cdot Z$, $Y = y \cdot
    Z$, and $T = X \cdot Y / Z$
    \cite{hisil2008twisted}.

    CycloneMSM \cite{aasaraai2022fpga} employs mixed PADD, which assumes two points
    are in different coordinate systems, such as affine and projective, thereby reducing
    the number of required multiplications. Similar to CycloneMSM, the present study
    utilized a mixed PADD in bucket accumulation and a complete PADD in bucket aggregation,
    as shown in Fig.~\ref{fig:padd}. To ensure symmetry in multiplication
    dataflow, we assumed that input points are provided in coordinates $(X, Y, U,
    Z)$, where $U = k\cdot T$ and $Z = 1$, and points in the bucket are represented
    in $(X, Y, T, Z)$. We extended the mixed PADD unit to a complete PADD for
    bucket aggregation by precomputing $U = k\cdot T$.

    \section{Proposal: Point Additions on AIE}
    \label{chap:method}
    \subsection{Overview Design}
    \label{sec:algorithm-overview} First, we present the overall algorithm for mixed
    PADD. We utilized the schoolbook algorithm and Barrett reduction for modular
    multiplications at the arithmetic level, as detailed in Sections~\ref{sec:mul}
    and \ref{sec:modmul-aie}. At the system level, the PL handles data transfers
    to and from DDR DRAM, initial additions, and final Barrett reductions. The
    AIEs perform multiplications and intermediate additions, with the optimizations
    discussed in Sections~\ref{sec:carry} and \ref{sec:padd}. Sections
    \ref{sec:fine} to \ref{sec:med} describe four strategies for the task
    partitioning and spatial mapping of PADD operations. Each strategy details how
    the schoolbook blocks are mapped to AIEs and how the overall PADD unit is organized
    within the AIE array.

    \subsection{Multiplication}
    \label{sec:mul}
    \begin{figure*}[htb]
        \begin{minipage}{0.37\textwidth}
            \centering
            \includegraphics[width=\textwidth]{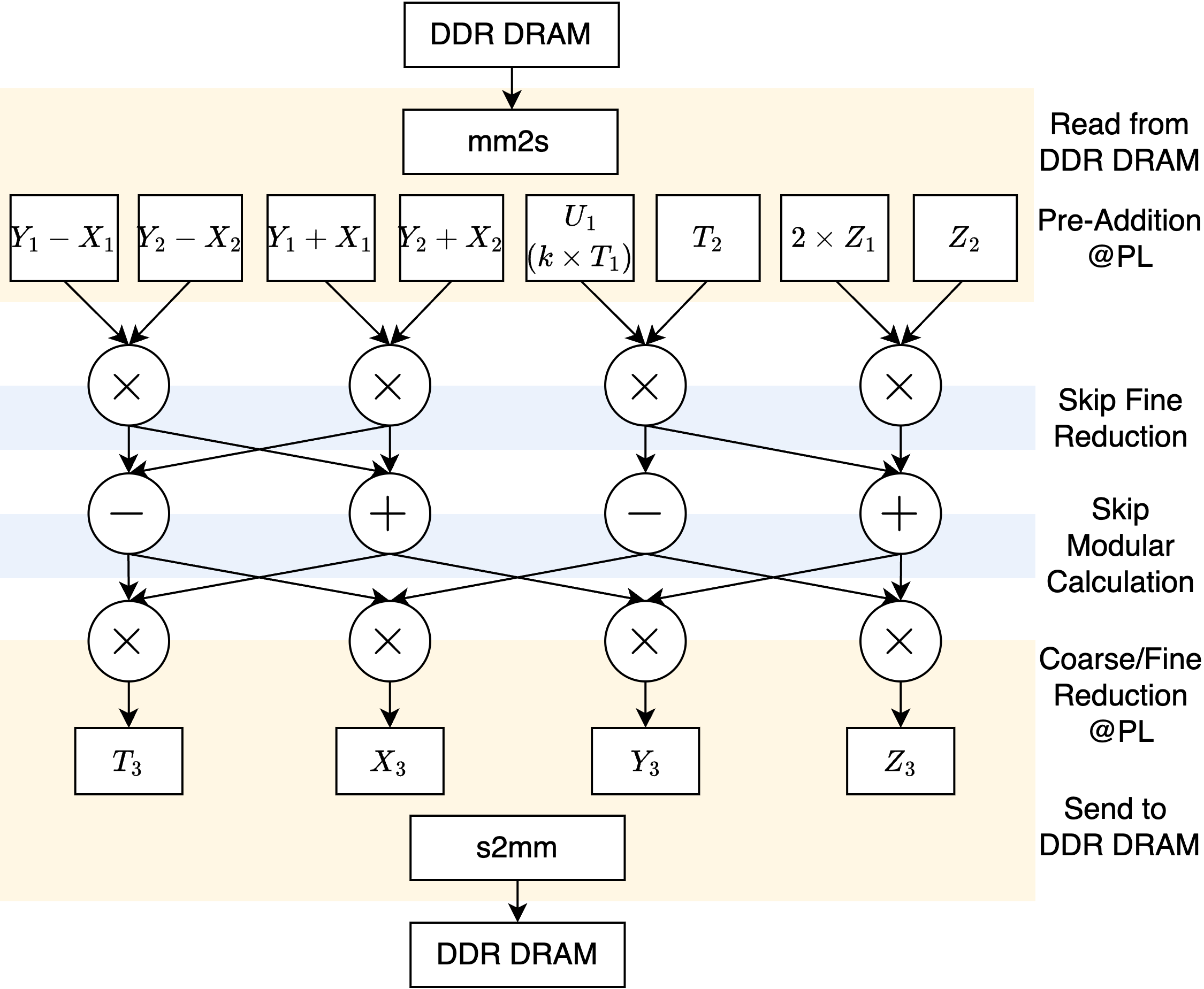}
            \caption{PADD operations}
            \label{fig:padd}
        \end{minipage}
        \hfill
        \begin{minipage}{0.62\textwidth}
            \centering
            \includegraphics[width=\textwidth]{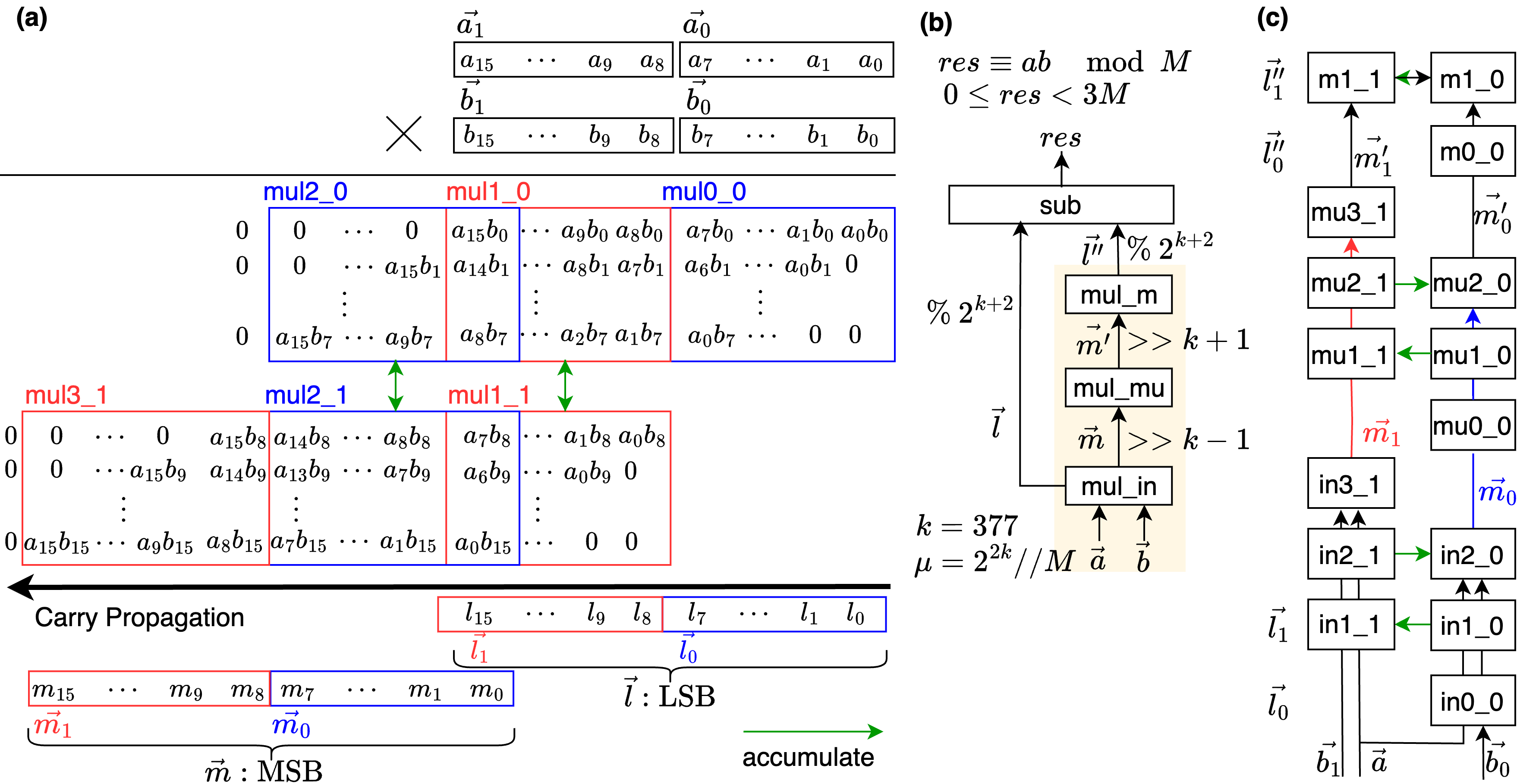}
            \caption{(a) Schoolbook algorithm and block partitions; (b) Overview
            of Barrett reduction; (c) Combination of the schoolbook algorithm
            and multiplications in Barrett reduction.}
            \label{fig:schoolbook}
        \end{minipage}
    \end{figure*}

    We employed a similar approach to AIM \cite{yang2023aim} for large integer
    representation and multiplication. We divided a 377-bit integer $a$ into 16 elements
    $a_{15}\cdots a_{1}a_{0}$; each 25-bit wide. We packed eight elements into a
    vector, representing each integer with two vectors
    $\{\vec{a_0}, \vec{a_1}\}$. We constructed a schoolbook algorithm and separated
    the computations into six blocks, as shown in Fig.~\ref{fig:schoolbook}(a). Each
    block processes eight 32-bit multiplications in a SIMD manner along the horizontal
    direction and then accumulates results along the vertical direction. For
    $\vec{b}$, blocks in the first row depend on $\vec{b_0}$, whereas those in
    the second row depend on $\vec{b_1}$. Each block involves data selection from
    $\vec{a}$ and either $\vec{b_0}$ or $\vec{b_1}$, multiplication, and accumulation,
    making it well-suited for the vector MAC operations of AIEs.
    \subsection{Modular Multiplication}
    \label{sec:modmul-aie} We adopted Barrett reduction for modular multiplication.
    This approach requires three multiplications, as shown in Fig.~\ref{fig:schoolbook}(b).
    First, we calculate the product of two inputs in \texttt{mul\_in}. Second,
    we multiply the MSB of the result by the constant $\mu$ in \texttt{mul\_mu}.
    Third, we multiply the MSB of \texttt{mul\_mu} by the modulus $M$ in \texttt{mul\_m}.
    Finally, we subtract the LSB of \texttt{mul\_m} from the LSB of \texttt{mul\_in},
    which results in the range of $[0, 3M)$. The constants $\mu$ and $M$ are assigned
    to $\vec{a}$ in Fig.~\ref{fig:schoolbook}(a).
    Fig.~\ref{fig:schoolbook}(c) illustrates how the blocks of the schoolbook algorithm
    map to Barrett reduction by aligning block partitioning with the inherent data
    dependencies. Specifically, the MSB of \texttt{mul\_in}, $\vec{m}= (\vec{m_0}
    , \vec{m_1})$, is passed to subsequent \texttt{mul\_mu} operations. In this
    process, \texttt{mu0\_0}, \texttt{mu1\_0}, and \texttt{mu2\_0} take inputs
    exclusively from $\vec{m_0}$, the output of \texttt{in2\_0}. Similarly,
    \texttt{mu0\_1}, \texttt{mu1\_1}, and \texttt{mu2\_1} rely solely on $\vec{m_1}$,
    the output of \texttt{in3\_1}. Consequently, the inputs for \texttt{mul\_mu}
    and \texttt{mul\_m} are drawn from a single block, enabling efficient tiling
    and minimizing data transfers.
    \subsection{Carry Propagation}
    \label{sec:carry} 
\begin{figure}[htb]
        \centering
        \includegraphics[width=0.48\textwidth]{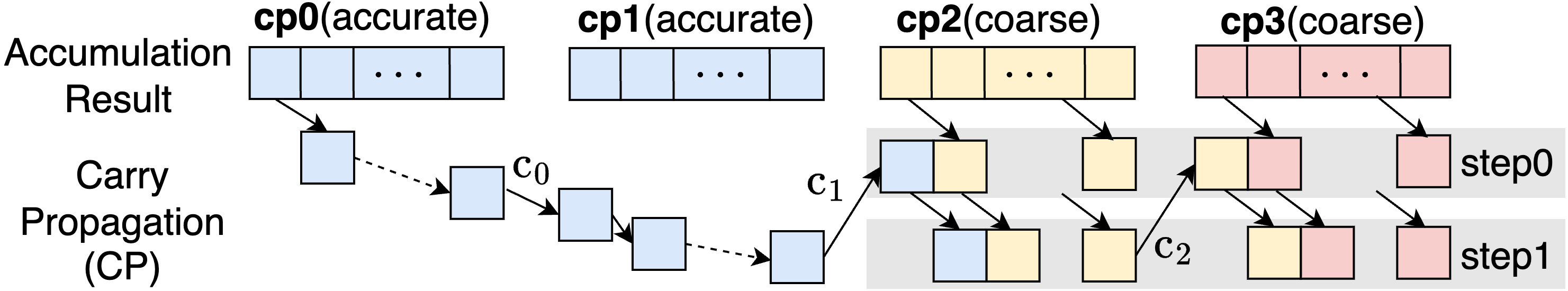}
        \caption{Carry propagation strategy. Regions with the same color are dependent.}
        \label{fig:carry}
    \end{figure}

Fig.~\ref{fig:carry} illustrates the carry propagation
    strategy. \texttt{cp0} represents the carry propagation after \texttt{mul0\_0},
    \texttt{cp1} occurs after \texttt{mul1\_0} and \texttt{mul1\_1}, and \texttt{cp2}
    and \texttt{cp3} follow the same pattern. We define two strategies: \textbf{accurate
    carry propagation} for \texttt{cp0} and \texttt{cp1} and \textbf{coarse
    carry propagation} for \texttt{cp2} and \texttt{cp3}.

    Accurate carry propagation is utilized when precise determination of
    successive MSB values is necessary. Carry bits are computed from the first element
    of the accumulated vector and propagated to the subsequent elements.
    Although this approach incurs numerous scalar operations, if accurate propagation
    is infrequent, the scalar operations can run in parallel with other vector
    operations through VLIW.

    Coarse carry propagation is employed when no exact magnitude comparison is
    needed, and keeping elements within the no-overflow range suffices. In our
    implementation, each element of the operand vector can hold up to 26 bits despite
    being separated by 25 bits. Coarse carry propagation repeats the following
    sequence twice (\texttt{step0} and \texttt{step1} in Fig.~\ref{fig:carry}):
    \begin{align*}
        \vec{\text{o}} & = \vec{\text{in}}\: \& \: \text{0x1ffffff} \\
        \vec{\text{c}} & = \vec{\text{in}}\gg 25                    \\
        \vec{\text{c}} & = [c_{\text{in}}] + \vec{\text{c}}[1:7]    \\
        \vec{\text{o}} & = \vec{\text{o}}+ \vec{\text{c}}
    \end{align*}
    Here, $\vec{\text{in}}$ denotes the accumulation result in \texttt{step0}, and
    the output $\vec{\text{o}}$ from \texttt{step0} becomes the input to \texttt{step1}.
    The carry $c_{\text{in}}$ comes from the previous block in \texttt{step0} and
    is set to 0 in \texttt{step1}. By utilizing the formula $x \: \& \: \text{0x1ffffff}
    = x - (x \gg 25) \ll 25$, we implement this entire sequence only with vector
    instructions, even though AIEs lack vector bitwise operations.

    Furthermore, coarse carry propagation increases parallelism between blocks
    by providing greater independence. While \texttt{cp0} and \texttt{cp1} depend
    fully on earlier computations, \texttt{cp2} is only partially dependent on the
    previous block, with its first two elements influenced by \texttt{cp0} and \texttt{cp1}.
    Similarly, \texttt{cp3} depends solely on \texttt{cp2}, again only for its first
    two elements.

    \subsection{PADD Optimizations}
    \label{sec:padd} We introduced two additional optimizations for PADD:
    omitting fine reduction from $[0, 3M)$ to $[0, M)$ in initial
    multiplications and skipping modular reduction in intermediate additions and
    subtractions, as shown in Fig.~\ref{fig:padd}. To prevent negative values in
    subsequent multiplications, we add $3M$ to the subtraction results. With these
    optimizations, inputs to final multiplications remain within the $[0, 6M)$
    range. Despite the expanded range, Barrett reduction remains effective by
    allowing coarse reduction within $[0, 4M)$, achieved through a looser reduction
    \cite{langhammer2021efficient} with a parameter $k$ set to 378.

    \subsection{Fine-grained Parallelism}
    \label{sec:fine} 
\begin{figure}[htb]
        \centering
        \includegraphics[width=0.49\textwidth]{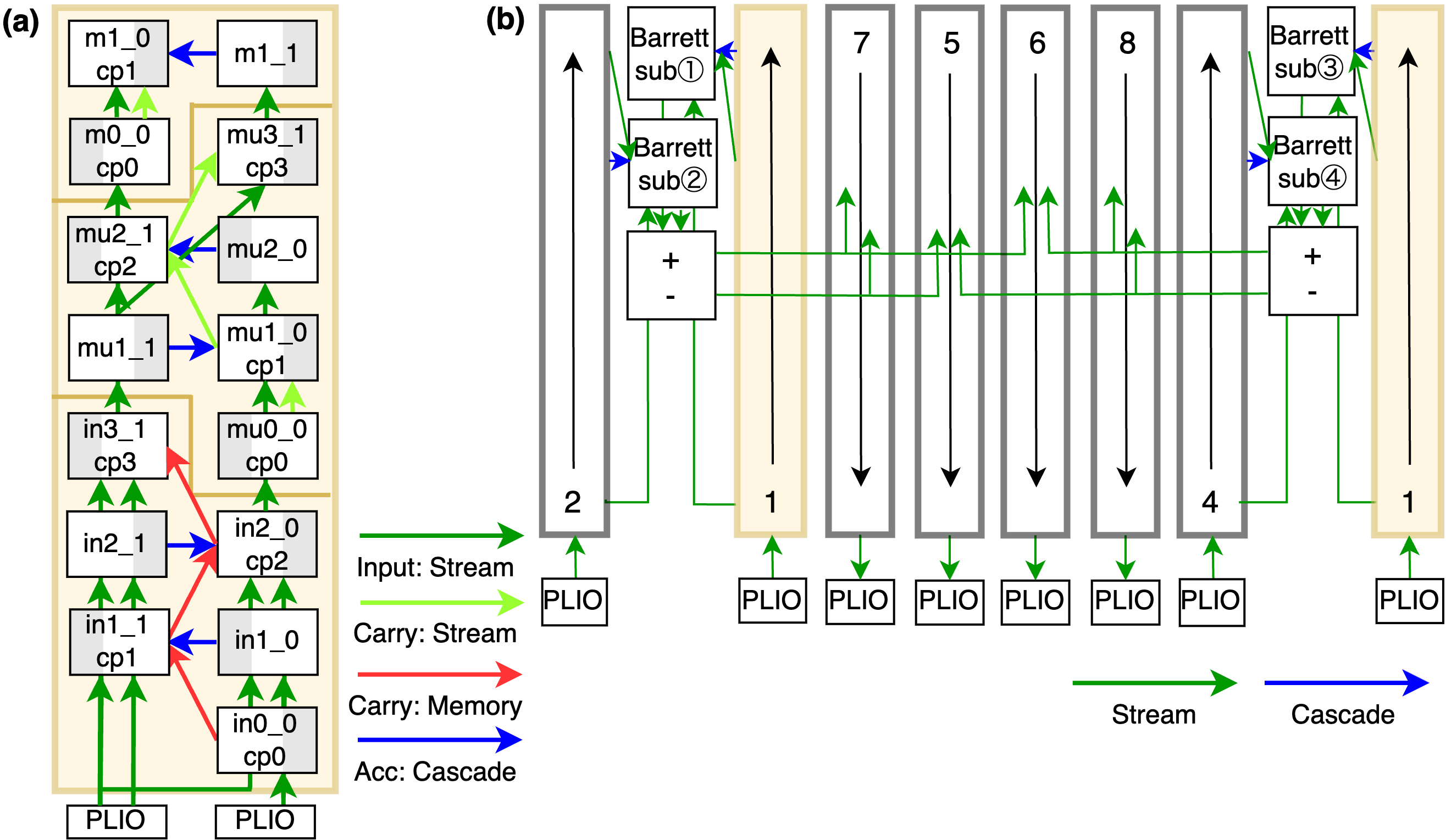}
        \caption{Fine-grained parallelism: (a) Mapping of Barrett reduction multiplications;
        (b) Mapping of overall PADD operations.}
        \label{fig:fine}
    \end{figure}

In the fine-grained parallelism, we assign each AIE tile to
    one block of the schoolbook algorithm. The main challenge lies in data
    movement between tiles. To address this, we propose optimized tiling and
    data movement strategies, illustrated in Fig.~\ref{fig:fine}(a). We use
    streams with broadcast for multiplication inputs, enabling data reuse across
    multiple blocks. Tiles are arranged to maximize data broadcast within the
    same column. For accumulated results, such as transfers from \texttt{in1\_0}
    to \texttt{in1\_1}, we employ cascade streams to manage large data transfers
    at high bandwidth. When stream connections are insufficient—such as for
    carry propagation in \texttt{mul\_in}—we use shared memory between
    neighboring tiles.

    Fig.~\ref{fig:fine}(b) illustrates the overall mapping of PADD. The first set
    of multiplications is tiled from rows 0 to 7, whereas the second set is
    tiled from rows 7 to 0.

    \subsection{Coarse-grained Parallelism}
    \label{sec:coarse} 
\begin{figure}[htb]
        \centering
        \includegraphics[width=0.30 \textwidth]{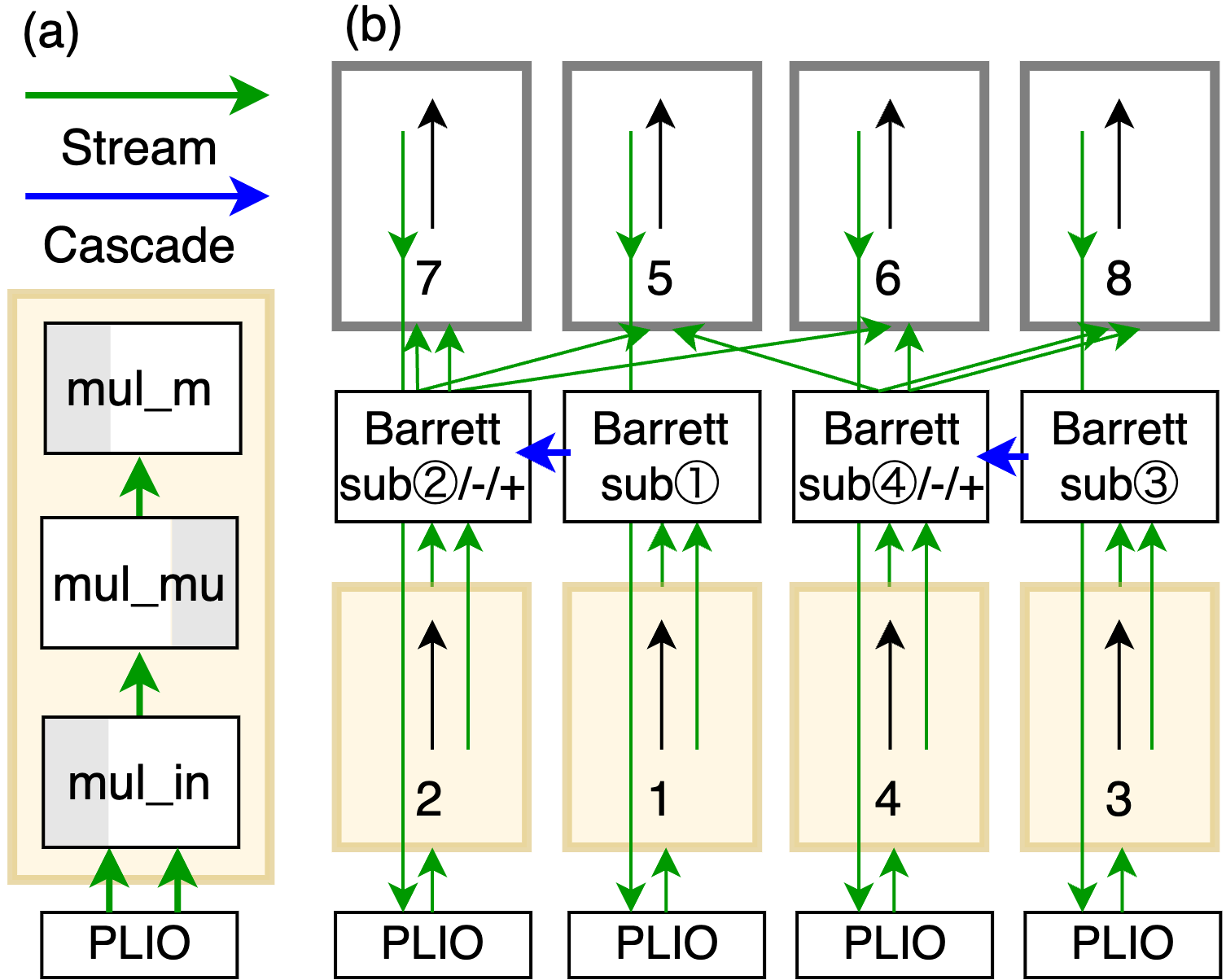}
        \caption{Coarse-grained parallelism: (a) Mapping of Barrett reduction multiplications;
        (b) Mapping of overall PADD operations}
        \label{fig:coarse}
    \end{figure}

In coarse-grained parallelism, we assign \texttt{mul\_in},
    \texttt{mul\_mu}, and \texttt{mul\_m} to each AIE core, as shown in Fig.~\ref{fig:coarse}(a).
    This approach has small data dependencies between tiles, and all data movement
    is handled without shared memory. Fig.~\ref{fig:coarse}(b) shows the overall
    design of PADD. The final set of modular multiplications is tiled on the top
    of the first set, and overall units require only four columns of the AIE
    array.

    \subsection{Medium-grained Parallelism}
    \label{sec:med} In medium-grained parallelism, we propose two strategies: cooperative
    (\textbf{med-coop}) and independent (\textbf{med-ind}), which are different
    by partitioning the schoolbook blocks.

    \subsubsection{Cooperative Strategy}

    We assign \texttt{mul0\_0}, \texttt{mul1\_0}, and \texttt{mul2\_0} to core A,
    and the remaining to core B, as shown in Fig.~\ref{fig:medium}(a)(b).
    Cascade streams connect the two cores to transfer accumulated results and
    the first block's carry. By processing distinct inputs—core A handling $\vec{m_{0}}$
    and core B handling $\vec{m_{1}}$—communication overhead via the small-width
    stream is reduced. Additionally, workload balance can be maintained without
    additional overhead by aligning the cycle counts across different cores. For
    instance, the second step of \texttt{cp2}, which could not be completed in
    \texttt{mul\_in B}, can be offloaded to \texttt{mul\_mu A}, which has fewer
    cycle counts.
    \subsubsection{Independent Strategy}

\begin{figure}[htb]
        \centering
        \includegraphics[width=0.49\textwidth]{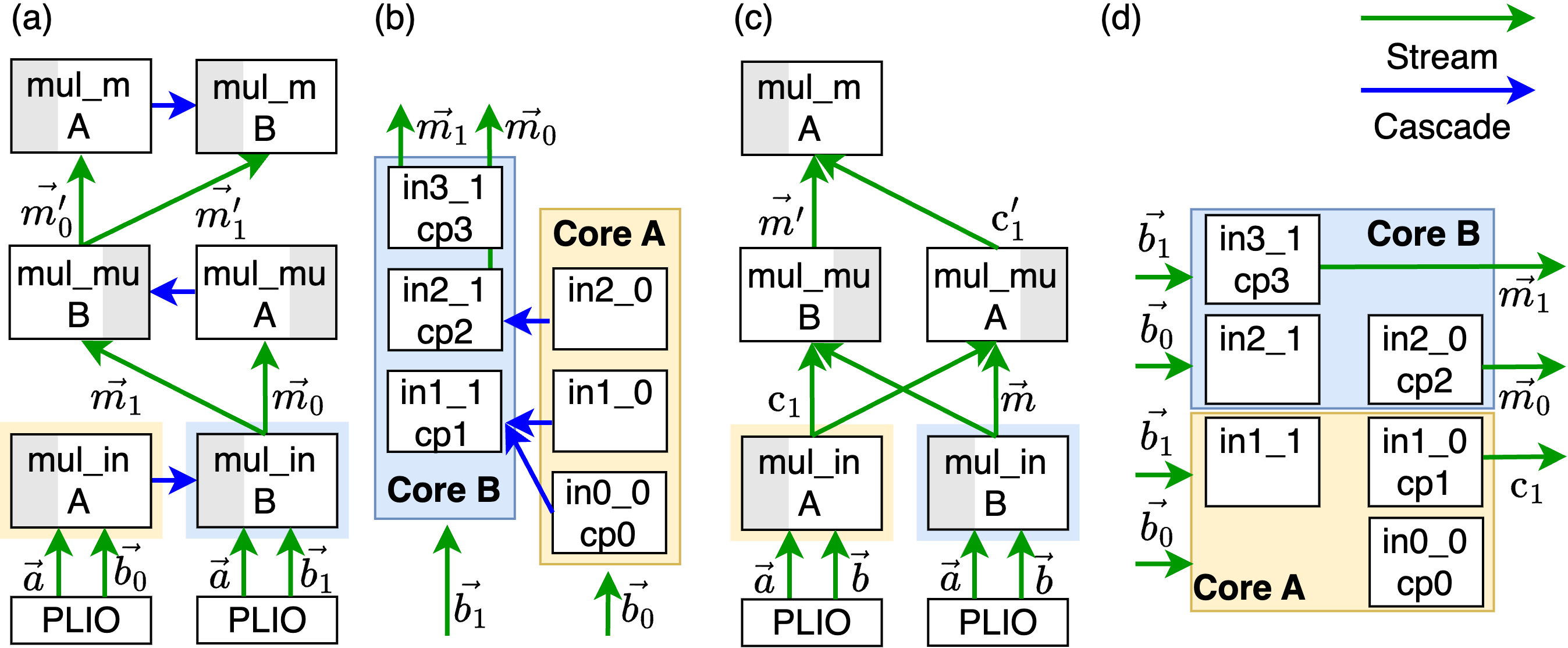}
        \caption{Medium-grained parallelism: (a)(b) cooperative strategy; (c)(d)
        independent strategy. (a)(c) Mapping of Barrett reduction
        multiplications; (b)(d) Division of schoolbook blocks into two cores.}
        \label{fig:medium}
    \end{figure}

    We assign \texttt{mul0\_0}, \texttt{mul1\_0}, and \texttt{mul1\_1} to core A
    and the remaining to core B, as shown in Fig.~\ref{fig:medium}(c)(d). We
    achieved complete independence between the two cores by leveraging the
    separation between the blue region and the yellow and red regions in Fig.~\ref{fig:carry}.
    Specifically, the first two bits of \texttt{cp2}, which depend on \texttt{cp1},
    are computed in the subsequent core—within \texttt{mul\_mu A} and \texttt{B}
    rather than in \texttt{mul\_in B}.

    The overall design of the medium-grained parallelism is the same as that of
    the coarse-grained parallelism, but the number of columns of the AIE array doubles
    to eight.

    \section{Cycle-accurate Simulation of AIE kernels}
    \label{sec:sim}
    \subsection{Setup}
    \label{subsec:setup} We compared the performance of four spatial mappings described
    in Section~\ref{chap:method}. We also evaluated the single AIE core
    implementation for comparison.
    We utilized the \textit{aiesimulator} to evaluate the performance of the
    proposed AIE kernels.

    We evaluated the results from two perspectives: \textbf{throughput} and
    \textbf{latency}. For the throughput, we first identified the maximum instruction
    count among all tiles and divided it by the number of parallelizable units.
    Each kernel was implemented as a loop with 16 iterations with software-pipelining,
    and we counted the number of instructions executed 15 or 16 times. For the
    latency evaluation, each kernel was designed to process only one set of inputs.
    The cycle count was recorded from when the first kernel was initialized to
    when the last kernel finished its operations.

    We also obtained total instructions and operations for discussion. For the total
    instructions, we summed the instruction counts across all tiles. The total operations
    were counted by decomposing VLIW instructions into individual operations,
    excluding \texttt{NOP}. Furthermore, we calculate the core utilization as \textit{total-instruction}
    / \textit{tiles-per-unit} / \textit{maximum-instruction}, tile utilization as
    \textit{tiles-per-unit} × \textit{parallelizable-unit} / 400, and VLIW
    utilization as \textit{total-operations} / \textit{total-instructions}.

    \subsection{Results}
    \label{subsec:results} 
\begin{figure}[htb]
        \centering
        \includegraphics[width=0.38\textwidth]{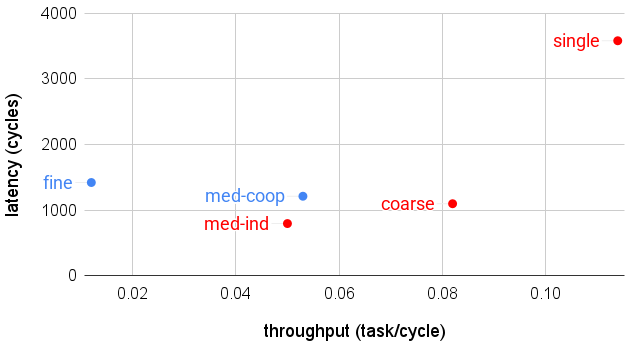}
        \caption{Latency and throughput for different configurations: Red points
        indicate the optimal candidates with high throughput and low latency.}
        \label{fig:chart}
    \end{figure}

Table~\ref{tab:cycle} summarizes the simulation
    results of the proposed AIE kernels. The \textit{single} achieved the
    highest throughput at 0.114 tasks per cycle, and the \textit{coarse}
    demonstrated the second-highest throughput at 0.082 tasks per cycle. Regarding
    latency, the \textit{med-ind} achieved the lowest value of 795 cycles,
    followed by the \textit{coarse} with 1097 cycles. As illustrated in Fig.~\ref{fig:chart},
    the \textit{single}, \textit{coarse}, and \textit{med-ind} are optimal candidates,
    offering high throughput and low latency.
    \begin{table}[htb]
        \centering
        \caption{Simulation results for different configurations.}
        \begin{tabular}{c | ccccc}
            \hline
            ~                   & fine  & med-coop & med-ind & coarse & single \\
            \hline
            columns             & 18    & 8        & 8       & 4      & 1      \\
            throughput(/cycle)↑ & 0.012 & 0.053    & 0.050   & 0.082  & 0.114  \\
            latency(cycle)↓     & 1420  & 1211     & 795     & 1097   & 3579   \\
            total instructions↓ & 8818  & 3940     & 3874    & 3466   & 3495   \\
            core utilization↑   & 0.42  & 0.80     & 0.88    & 0.85   & 1.00   \\
            tile utilization↑   & 0.63  & 0.78     & 0.66    & 0.84   & 1.00   \\
            VLIW utilization↑   & 0.88  & 1.76     & 1.73    & 1.98   & 1.85   \\
            \hline
        \end{tabular}
        \label{tab:cycle}
    \end{table}
    \subsection{Discussion}
    \label{subsec:discussion}
    \subsubsection{Throughput}
    Increasing the number of tile partitions reduces throughput due to four key factors,
    as shown in Table~\ref{tab:cycle}. First, finer partitioning increases
    communication costs, leading to a higher total instruction count. In the
    \textit{fine}, memory access is used for carry propagation to compensate for
    the lack of stream connections; however, acquiring a lock for each access
    introduces 20 to 30 NOP instructions. Second, the workload distribution
    across tiles becomes unbalanced, resulting in stalls in faster tiles,
    leading to low core utilization. Third, finer partitioning reduces resource
    utilization by limiting inter-task parallelism per unit. The \textit{fine}
    supports only two parallel instances, leaving 16 columns unused. Lastly, smaller
    coherent code blocks lead to lower VLIW utilization. In contrast to the
    \textit{coarse}, where element-wise operations are combined with MAC
    operations, the \textit{fine} executes MAC operations separately.

    \subsubsection{Latency}
\begin{figure}[htb]
        \centering
        \includegraphics[width=0.49\textwidth]{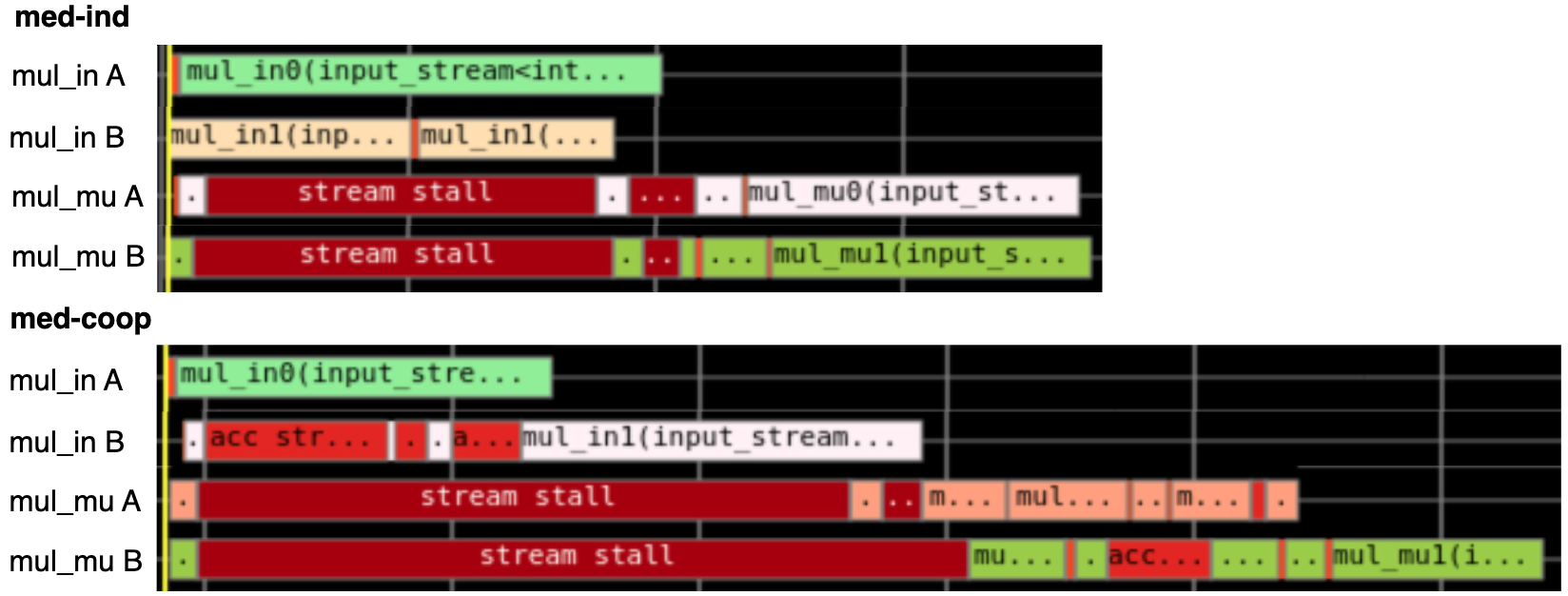}
        \caption{Timeline trace of \textit{med-ind} and \textit{med-coop}, with the
        horizontal axis representing time and all blocks shown on the same scale.
        Red blocks indicate stalls, while others represent computations.}
        \label{fig:latency-trace}
    \end{figure}

    Fig.~\ref{fig:latency-trace} illustrates the timeline trace of the \textit{med-ind}
    and \textit{med-coop}. The \texttt{mul\_in A B} and \texttt{mul\_mu A B}
    blocks correspond to the tiles in Fig.~\ref{fig:medium}. The latency of
    \textit{med-ind} was 0.65 times lower than that of \textit{med-coop}, although
    they have approximately the same instruction count. This difference arises
    because, in \textit{med-ind}, the processing tasks of cores A and B are independent,
    allowing them to execute simultaneously. Conversely, in \textit{med-coop},
    the results from core A are cascaded to B, and the compiler schedules operations
    of core B, even those unrelated to core A, only after reading the results of
    core A. Implementing finer-grained function blocks could encourage the
    compiler to schedule cores A and B more cooperatively. However, this would
    reduce the range of interchangeable code sections and lower the VLIW
    utilization. Thus, broadly structuring independent processing blocks and
    assigning separate cores reduces latency.

    \subsubsection{Carry Propagation Overhead}
    Our evaluation shows that integrating carry propagation within the AIE
    reduces communication overhead and improves efficiency. If carry propagation
    is handled in the PL instead, the AIE must send each 377-bit multiplication
    result back, which requires transferring 42.6 bits per cycle. Specifically, each
    multiplication involves 48 MAC operations and returns 32 64-bit elements,
    leading to a bandwidth requirement of $32 \times 64 \div 48$. With the PL operating
    at 312.5 MHz and a PLIO bandwidth of 4.68 Tb/s, each of the 400
    tiles—running at 1.25 GHz—can transmit only 9.36 bits per cycle. This meets just
    21.9\% of the required bandwidth and caps MAC utilization at 21.9\%. In
    contrast, performing carry propagation within the AIE requires 3495 cycles,
    including 960 MAC operations, resulting in a 27.4\% MAC utilization—1.25$\times$
    higher than the PL-based approach.

    \section{Experiments on Real Platform}
    \label{sec:real}
    \subsection{Setup}
    \label{subsec:setup-real}

    From the configurations compared in Section~\ref{sec:sim}, we selected
    \textit{coarse} for bucket accumulation, which achieved the second-highest
    throughput. The \textit{single} approach, with the highest throughput,
    requires packet-switching to independently distribute data to cores, incurring
    additional implementation and scheduling costs. Moreover, a lower latency helps
    reduce BRAM usage when scheduling successive inputs. We chose \textit{med-ind}
    for bucket aggregation, which has the smallest latency. We integrated them with
    PL and PS components for deployment on the VCK190 platform. The PL modules
    were implemented using Vivado HLS and operate at 312.5 MHz.

    We evaluated the throughput with 10-way parallelism, allowing simultaneous mapping
    of the \textit{med-ind} design. The PL kernel accepts one input point $(X, Y,
    U)$, with the second operand replaced by a dummy point, and returns the result
    only in the final cycle. This design represents the behavior of MSM bucket
    accumulation. We measured the time to complete the addition of $2^{18}$ points.

    We evaluated the latency with complete PADD by multiplying the constant $k$ with
    the input $T_{1}$ before performing mixed PADD. The $k$-multiplication was
    mapped in row 0 of the columns already used for mixed PADD. The PL accepts one
    input point, adds it to the previous accumulated value, and processes the next
    input point only after completing the previous addition. The latency is
    measured as the time to complete the sum of $2^{18}$ points.

    We compared the throughput and latency of the proposed AIE kernels with
    those of the BSTMSM \cite{yang2023aim} and CPU implementations. We estimated
    its throughput and latency based on the reported PADD units and cycle counts.
    BSTMSM includes two PADD units and operates as a fully pipelined system with
    a cycle count of 260 cycles at 300 MHz. For the CPU implementation, we
    utilized GMP Libary \cite{gmp} for large integer arithmetic, and measured on
    the CPU of the VCK190 platform.

    \setlength{\textfloatsep}{4.5pt}
    \setlength{\intextsep}{4.5pt}

    \begin{table*}
        [t]
        \centering
        \caption{Throughput comparison of the proposed approach with existing
        accelerators. ${}^{*}$ denotes the estimated throughput.}
        \begin{tabular}{c|cccccc}
            \hline
            ~                     & BSTMSM${}^{*}$ \cite{zhao2023bstmsm} & CPU    & PL ${}^{*}$ & Ours   & Ours + PL ${}^{*}$ & Ours + PL ${}^{*}$ \\
            \hline
            device                & U250                                 & VCK190 & VCK190      & VCK190 & VCK190             & VCK190             \\
            configuration         & -                                    & -      & -       & coarse      & coarse             & single             \\
            parallelism           & 2                                    & -      & 1          & 10      & 12                 & 400                \\
            \hline
            throughput(M task/s)↑ & 600.0                                & 0.11   &   300.0    & 67.0    & 403.4              & 443.0              \\
            \hline
        \end{tabular}
        \label{tab:th-cmp}
    \end{table*}

    \subsection{Results}
    \label{subsec:results-real}

    \subsubsection{Throughput}
    Table~\ref{tab:th-cmp} compares the throughput of the proposed approach with
    the BSTMSM and CPU. The proposed system achieved a throughput of 67.0 M task/s,
    utilizing 50.2\% of the DRAM bandwidth, 568 times higher than the CPU
    implementation. However, its performance is only one-ninth of that achieved
    by BSTMSM. To explore potential performance benefits beyond the PL-only implementation,
    we consider a scenario where the PL also performs PADD alongside the AIE. We
    assume that the PL in VCK190 accommodates a single PADD unit, the one used
    in BSTMSM, and the AIE operates with 12 parallel coarse units or 400 parallel
    single units, as simulated in Section~\ref{sec:sim}. Under this assumption,
    it achieves a 1.34$\times$ or 1.47$\times$ speedup over the PL-only approach,
    demonstrating the potential advantages of integrating the AIE with PL resources.
    Note that bandwidth limitations in this scenario can be addressed by
    utilizing additional external memory of LPDDR DRAM.

    \subsubsection{Latency}
    Table~\ref{tab:latency} compares latency and resource utilization. The proposed
    system achieved a latency of 1.05 µs, significantly lower than the CPU. In contrast,
    the proposed system's latency is 1.22 times higher than BSTMSM. However, the
    proposed system achieves the observed latency using only 0.28 times the
    multiplier resources BSTMSM requires when converted to INT32 multipliers. The
    AIE's independent and parallel processing capabilities were leveraged effectively
    to maintain performance while minimizing resource usage.
    \begin{table}[!ht]
        \centering
        \caption{Comparison of latency and resources.}
        \begin{tabular}{c|ccc}
            \hline
            ~                     & BSTMSM \cite{zhao2023bstmsm} & CPU  & Ours \\
            \hline
            Latency($\mu$s/task)↓ & 0.86                         & 9.69 & 1.05 \\
            DSP48E2s              & 2920                         & -    & 0    \\
            AIE Tiles             & 0                            & -    & 50   \\
            INT32 Multipliers     & 1385.8                       & -    & 400  \\
            \hline
        \end{tabular}
        \label{tab:latency}
    \end{table}

    \subsection{Discussion}
    \label{subsec:discussion-real} The proposed system underperforms BSTMSM in throughput
    due to two primary factors.

    First, whereas FPGAs benefit from dedicated carry logic (\textit{CARRY8}),
    AIE lacks a similar feature and must rely on the scalar unit, which processes
    only one arithmetic per cycle. This architectural limitation constrains MAC
    utilization despite the capability to perform eight multiplications per
    cycle. Carry propagation could be more efficient if the AIE supported finer-grained
    control over scalar operations. As a potential algorithmic improvement, adopting
    Montgomery representation will reduce the need for accurate carry propagation,
    mitigating reliance on scalar operations.

    Second, the multiplication requirement is five times higher. SIMD processing
    wastes 66\% of multiplications on constant zero values. Additionally, BSTMSM
    leverages the Karatsuba algorithm, which has a time complexity $O(N^{1.58})$.
    In contrast, AIE relies on the less efficient schoolbook method with the
    complexity of $O(N^{2})$, owning to Karatsuba's poor data reusability. A
    more balanced approach—using Karatsuba at a high level while retaining the
    schoolbook at a low level—could enhance data reusability and computational
    efficiency.
    \section{Conclusion}
    In this study, we proposed PADD acceleration on the Versal AI Engine for MSM.
    Unlike existing large integer multiplication accelerators, we introduced an
    AIE-centric approach leveraging VLIW and SIMD capabilities. Additionally, we
    proposed four distinct spatial mapping strategies and compared their
    throughput and latency. The proposed AIE kernels achieved a throughput of 67.0
    M task/s, 568 times higher than the CPU implementation. We also analyzed the
    efficiency of the proposed algorithm and confirmed that performing carry propagation
    on the AIE, rather than on the PL, improves MAC utilization. Furthermore, we
    suggested potential improvements through comparisons with FPGA-based accelerators.


    \section*{Acknowledgment}
    This work is supported in part by JST CREST JPMJCR21D2.

    \renewcommand{\baselinestretch}{0.95}


\end{document}